\documentclass[prb,twocolumn,superscriptaddress,floatfix]{revtex4}
\usepackage{hyperref}
\usepackage{graphicx} 
\usepackage{amsmath}
\usepackage[dvips]{epsfig}
\newcommand{\beq}{\begin{equation}}
\newcommand{\eeq}{\end{equation}} 
\newcommand{\beqa}{\begin{eqnarray}}
\newcommand{\eeqa}{\end{eqnarray}}
\newcommand{\ba}{\begin{array}}
\newcommand{\ea}{\end{array}}

\begin{document}

\title{Spinning superfluid helium-4 nanodroplets}

\author{Francesco Ancilotto}
\affiliation{Dipartimento di Fisica e Astronomia ``Galileo Galilei''
and CNISM, Universit\`a di Padova, via Marzolo 8, 35122 Padova, Italy}
\affiliation{ CNR-IOM Democritos, via Bonomea, 265 - 34136 Trieste, Italy }

\author{Manuel Barranco}

\affiliation{Departament FQA, Facultat de F\'{\i}sica,
Universitat de Barcelona. Diagonal 645,
08028 Barcelona, Spain}
\affiliation{Institute of Nanoscience and Nanotechnology (IN2UB),
Universitat de Barcelona, Barcelona, Spain.}
\affiliation{Universit\'e Toulouse 3 and CNRS, Laboratoire des Collisions, Agr\'egats et R\'eactivit\'e,
IRSAMC, 118 route de Narbonne, F-31062 Toulouse Cedex 09, France
}

\author{Mart\'{\i} Pi}
\affiliation{Departament FQA, Facultat de F\'{\i}sica,
Universitat de Barcelona. Diagonal 645,
08028 Barcelona, Spain}
\affiliation{Institute of Nanoscience and Nanotechnology (IN2UB),
Universitat de Barcelona, Barcelona, Spain.}

\begin{abstract} 
We have studied spinning superfluid $^4$He nanodroplets at zero temperature
using Density Functional theory.
Due to the irrotational character of the superfluid flow,
the shapes of the spinning nanodroplets are very different
from those of a viscous normal  fluid drop in steady rotation.
We show that when vortices are nucleated inside  the superfluid  droplets,
their morphology, which evolves from axisymmetric oblate to triaxial prolate to
two-lobed shapes, is in  good agreement with experiments.
The presence of vortex arrays confers to the superfluid droplets
the rigid-body behavior of a normal fluid in steady  rotation, and
this is the  ultimate reason of
the surprising good agreement between recent
experiments and the classical models used for their description.
\end{abstract} 
\date{\today}


\maketitle

\section{Introduction}

When a drop made of a normal liquid rotates,
centrifugal forces 
produce deformations which change its spherical appearance
at rest into an oblate, axisymmetric shape.
If the drop spins fast enough, it may eventually distort
into a two-lobed, peanut-shaped form before undergoing fission
when it is no longer able to sustain
the strain due to its own rotational motion.
Hydrodynamical models aiming at describing rotating fluid drops
have been successfully applied to systems that 
range from atomic nuclei to celestial objects. 
Most theoretical models assume the liquid to be 
incompressible and viscous so that drops eventually
reach a steady  state in which they rotate as if they were rigid bodies.
\cite{Cha65,Coh74,Bro80,Bro80b,Nur15,Hei06,Hil08,But11,Bal15,Lia17}

Using wax samples under diamagnetic levitation, the 
theoretical shapes of  rotating drops were 
reproduced experimentally.\cite{Bal15}
This study  has provided
direct experimental validation of numerical 
models used to calculate equilibrium shapes 
of spinning drops made of classical fluids.
With increasing angular momentum, the shapes 
of the wax samples progress 
from spheres (not rotating) to oblate-like shapes, to tri-axial 
shapes (three unequal axes), and finally to two-lobed (dumbbell) shapes. 
In the latter case the axis of rotation is perpendicular 
to the line joining the centers of the two lobes.
For all shapes, the shortest 
axis always coincides with the axis of rotation.

With the experimental realization of superfluid 
helium drops using cryogenic free-jet gas expansions,
\cite{Toe04} attention has been focused recently on the shapes of superfluid  drops.
It is commonly accepted 
that helium drops created in the normal, non-superfluid 
phase, may acquire angular momentum 
during the passage of the fluid through the nozzle 
of the experimental apparatus.\cite{Ber17}
Vorticity, defined in hydrodynamics as $\nabla \times \mathbf{v}$,\cite{Guy15} 
where $\mathbf{v}$ is the velocity field of the fluid,
is distributed inside the drop in the normal phase. It  
equals $2 \omega$ for a rotating rigid body, or for a 
viscous fluid in steady rotation,  $\omega$ being the angular velocity about the
rotation axis.

During the expansion process the helium drops cool 
down to  about 0.4 K temperature;
at this temperature, the helium is superfluid and 
the normal fluid fraction 
is negligible. The spinning  superfluid drops 
retain a large yet unknown 
fraction of the angular momentum deposited 
in them when they are still in the normal phase.
Since the superfluid flow is irrotational,
$\nabla \times \mathbf{v}=0$, the  vorticity remaining in the drops
is concentrated in the cores of the quantized vortices 
nucleated in the drop interior\cite{Don91,Fet74} and in 
capillary waves.\cite{Whi98,Fet74} 
This is the mechanism 
by which angular momentum is conserved
in an isolated helium drop undergoing a 
normal to superfluid state transition. 

 \begin{figure}[!]
\centerline{\includegraphics[width=1.0\linewidth,clip]{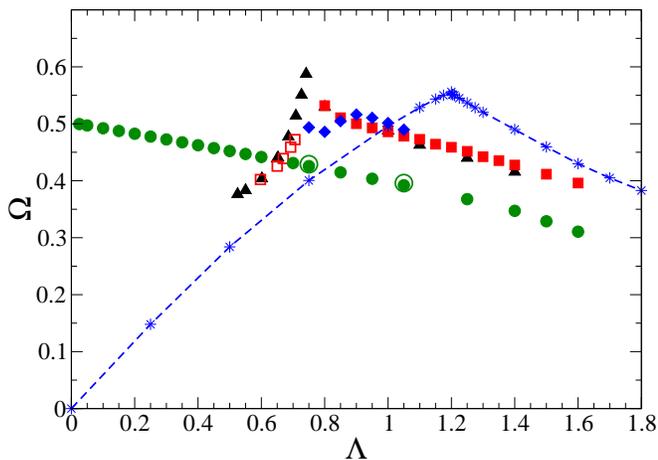}}
\caption{
Rescaled angular velocity $\Omega$ {\it vs.} rescaled angular momentum $\Lambda$.
Black triangles: 3-vortex configurations. Red squares: 4-vortex configurations (open, oblate-like; solid, linear).
Blue diamonds: 4-vortex cross configurations. Green dots: vortex-free configurations;
big empty green dots show values for the  $^4$He$_{5000}$  droplet.
The blue starred symbols  connected by a dashed line are the 
classical rotating drop results of Ref. \onlinecite{But11}.
}
\label{fig1}
\end{figure}

Superfluid $^4$He drops in fast rotation
have been studied by coherent x-ray scattering.\cite{Gom14}
The shapes of the drops observed  
were consistent with those of axisymmetric oblate pseudo-spheroids with large aspect
ratio, defined as the ratio of the long half-axis length to the
short half-axis length along the rotational axis.
The existence of vortex arrays
inside a number of  drops was  established by the appearance
of Bragg patterns from Xe clusters trapped in the vortex cores.\cite{Gom14,Jon16}
Doping the drops  was instrumental for 
detecting the presence of vortices,
the extremely small size of the vortex cores
preventing their direct detection. 
At variance, the shapes of the drops could be 
instead inferred in the absence of doping.

While normal fluid drops change their
shape as rotation becomes faster to resemble
a ``peanut'' (two-lobe shape) or a ``blood cell'',\cite{Cha65,Bro80,Hil08} no evidence of such
shape shifting was seen in the  
experiments  of Ref. \onlinecite{Gom14}.
Density Functional Theory (DFT) calculations carried out for
droplets containing 15000 $^4$He atoms and 
up to $n_v=9$ vortices,
confirmed the above scenario,\cite{Anc15,Cop17}
the comparison 
with the experimental measurements being facilitated 
by using suitably rescaled units to characterize the droplet shapes (see below), 
which showed indeed
remarkable similarities with the experimental drops.
Quite a  few helium droplets displaying prolate shapes were subsequently identified  in these    
small-angle, x-ray diffraction experiments.\cite{Ber17} 
So far, no theoretical analysis of the prolate 
configurations found in the experiments has  been carried out. 
This is one of the purposes of the present work.
 
Very recently,   coherent diffractive imaging experiments of rotating $^4$He drops using 
extreme ultraviolet pulses in conjunction with wide-angle x-ray diffraction  have  allowed to identify, in addition to
oblate shapes, a large number of prolate shapes.\cite{Rup17,Lan18}
These experiments have been analyzed  
by simulating the observed diffraction patterns 
obtained from a simple parametrization model 
for the drop geometry  (a combination of two ellipsoidal caps smoothly connected 
by a hyperboloidal centerpiece),  and comparing them with the actual 
diffraction patterns until a match is found.

Comparison of the parametrized shapes of the experimental drops
with those predicted by numerical calculations for normal liquid rotating drops\cite{Bal15}  
has shown
a good agreement,\cite{Lan18}
indicating that 
helium drops formed in the free jet expansion 
follow closely the sequence of shapes 
characteristic of  normal fluid drops.

The most natural question arises, i.e. why 
spinning superfluid $^4$He drops, whose  hydrodynamical 
hallmark is irrotational flow,
behave instead as rotating normal 
liquid drops. 
The answer, as we will quantitatively show in the following, 
is that
the presence of vortex arrays in the helium drops
makes them  behave as classical rotating liquid droplets.

The similarity between 
rotating superfluid $^4$He and the rigid-body rotation of
a viscous liquid has been noted long ago
for bulk $^4$He  in a  rotating bucket: in spite of its irrotational flow,
$^4$He develops  a meniscus just like a 
rotating normal fluid, instead of remaining at rest with a flat 
liquid-vapor interface.\cite{Osb50}
This apparently contradictory behavior has been attributed to
the nucleation of quantized vortices which
carry most of the angular momentum of the 
rotating fluid.\cite{Pit16} 

\section{Method}

We  have used Density Functional Theory  to 
describe  superfluid helium droplets.\cite{Anc17} Within DFT,
the total energy  of a $^4$He$_N$ droplet  
at zero temperature is written as a functional
of a complex effective wave function $\Psi( \mathbf{r})$
related to its atomic density by $\rho (\mathbf{r})= |\Psi( \mathbf{r})|^2$.

\begin{figure}[!]
\centerline{\includegraphics[width=0.7\linewidth,clip]{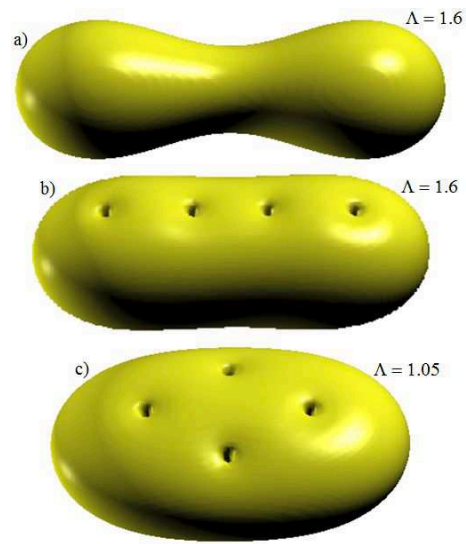}}
\caption{
Prolate $^4$He$_{1500}$ droplet shapes (not to scale) represented by their sharp density surfaces
and $\Lambda$ values.
a)  Vortex-free configuration; b) 
4-vortex linear configuration; c) 
4-vortex  cross configuration.
The rotational axis coincides with the vertical axis in the figure. 
}
\label{fig2}
\end{figure}
 
The droplet equilibrium configuration is obtained by solving the equation
\begin{equation}
\left\{-\frac{\hbar^2}{2m_{\rm He}} \nabla^2 + \frac{\delta {\cal E}_c}{\delta \rho}
\right\}\Psi({\mathbf r}) 
 \equiv
{\cal H}[\rho] \,\Psi({\mathbf r}) 
= \mu_4 \Psi({\mathbf r})
\label{eq2}
\end{equation}
where $\mu_4$ is the $^4$He chemical potential and ${\cal E}_c[\rho]$ 
is the correlation energy, for which we 
have taken that proposed in  Ref.  \onlinecite{Anc05}. 
This functional is finite-range and includes non-local effects.
Both aspects are needed to  
describe quantitatively the response of the liquid at the \AA-scale.
Unless explicitly stated, the number of helium atoms
in the  studied droplets is $N=1500$. 

To investigate non-zero angular momentum configurations in the spinning droplet
it is convenient to move to the fixed-droplet frame of 
reference (corotating frame)
by imposing, through the use of a Lagrange multiplier $\omega$, 
a fixed value for the total angular momentum $\langle L_z \rangle$, 
i.e. we look for solutions of the equation 
\begin{equation}
\{{\cal H}[\rho] \,-\omega \, \hbar \hat{L}_z\} \,\Psi(\mathbf{r})  =  \,\mu_4 \,
\Psi(\mathbf{r}) 
\label{eq3}
\end{equation}
where $\hat{L}_z$ is the dimensionless angular momentum operator.
The results presented in this work have been obtained 
using the 4He-DFT BCN-TLS computing package.\cite{Pi17}
Details on how Eqs. (\ref{eq2}) and (\ref{eq3})  
are solved can be found in  Ref. \onlinecite{Anc17} and references therein.
We work in 3D cartesian 
coordinates and  no symmetry is imposed to the
solution of Eq.(\ref{eq3}) during the functional minimization. 

\begin{figure}[!]
\centerline{\includegraphics[width=1.0\linewidth,clip]{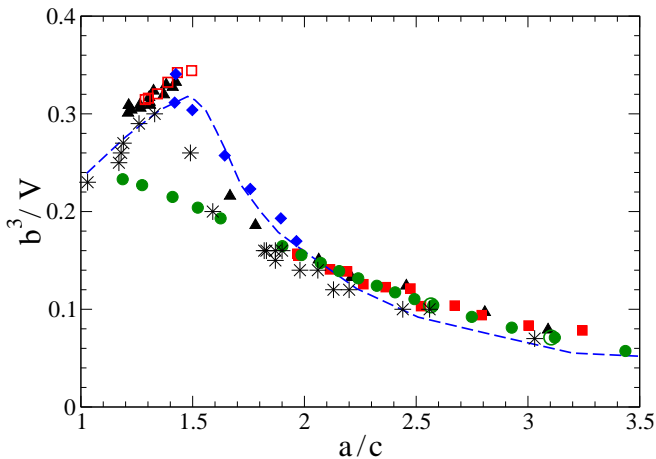}}
\caption{
Aspect-ratio  $b^3/V$ {\it vs.}   $a/c$ curve.
Black triangles: 3-vortex configurations. Red squares: 4-vortex configurations (open, oblate-like; solid, linear).
Blue diamonds: 4-vortex cross configurations. Green dots: vortex-free configurations;
big empty green dots show values for the  $^4$He$_{5000}$  droplet.
The black starred symbols show the experimental results of Ref. \onlinecite{Lan18}.
The blue dashed line shows the 
classical rotating drop results of Ref. \onlinecite{Bal15}.
}
\label{fig3}
\end{figure}

As in previous studies,\cite{Hei06,But11,Ber17} we  use  here 
rescaled units for 
angular momentum, $\Lambda$, and angular velocity, $\Omega$. This allows to compare our 
results, which are obtained for droplets made of $O(10^3)$ atoms,
with experimental results on large drops, made of $10^8-10^{11}$ atoms. 
Such units are defined as\cite{note}
\begin{eqnarray}
 \Omega &\equiv& \sqrt{\frac{m_{\rm He} \, \rho_0 \, R^3}{8 \, \gamma}}\; \; \omega 
 \\
\nonumber
\Lambda &\equiv&\frac{\hbar}{\sqrt{8 \gamma R^7  m_{\rm He}\rho_0}} \;\; L_z 
\label{eq4}
\end{eqnarray}
where $\gamma$ = 0.274 K \AA$^{-2}$ and 
$\rho_0$ = 0.0218 \AA$^{-3}$ are the surface tension and liquid 
atom density at zero temperature and pressure;
$R= 2.22\,N^{1/3}$ \AA{} is the sharp radius of the spherical 
droplet with $N$ helium atoms and zero angular momentum, 
defined such that $4\pi R^3 \rho_0 /3= N$.
Liquid helium is fairly incompressible and hence the volume of 
any deformed configurations can be safely identified with $V= 4 \pi R^3/3$. 

We have employed  two different, alternative strategies to solve  
Eq. (\ref{eq3}), i.e. we either  (i)  fix $\omega $ and 
find the associated stationary configuration,
which will be characterized by some value of the angular momentum $L_z$
depending upon the chosen value of $\omega$, or (ii) solve Eq. (\ref{eq3})
by imposing a given value for  $L_z$  and
iteratively find the associated value of $\omega$.
Classically, the fixed $\omega$ calculations correspond to 
forced rotation conditions (``driven drops''), while the fixed $L_z$ calculations correspond 
to torque free drops with an initial prescribed rotation (``isolated drops'').\cite{Bro80,Nur15} 
As shown in Ref. \onlinecite{Bro80},
it turns out that stable prolate configurations can only been found fixing the value of $L_z$.
At variance, stable
oblate configurations can be found either by fixing $\omega$ or $L_z$. 

\section{Results}

The $\Omega(\Lambda)$ relationship 
for  normal fluid drops in steady rotation is plotted 
with a dashed line in Fig. \ref{fig1}.\cite{But11} 
Configurations with $\Lambda \lesssim 1.2$ 
have oblate axisymmetric shapes, and configurations with 
larger $\Lambda$ values have  prolate triaxial or  
two-lobed shapes;\cite{Bro80, Bro80b,Hei06,But11}
the classical 
bifurcation point is at $(\Lambda, \Omega) \sim (1.2, 0.56)$.\cite{note}

As shown  also in Fig. \ref{fig1},  the calculated  $\Omega(\Lambda)$ relation for a 
superfluid droplet (filled dots) is very different. In this case,  
only prolate triaxial  configurations may exist since  
axisymmetric oblate configurations are  quantum mechanically forbidden.\cite{Vil57}
The finite value of $\Omega$ at very small values of $\Lambda$
is the equivalent of the ``rotational Meissner effect'' 
occurring when
liquid helium in a rotating cylinder is cooled
through the lambda point:
at sufficiently slow rotational
speeds the superfluid forms in a state of zero total
angular momentum, causing the container to rotate faster.\cite{Hes67}
A nearly spherical configuration with a high value of $\Omega$ 
and a negligible value of $\Lambda$ can be seen for instance 
in Fig. 8 of Ref. \onlinecite{Cop17}.

The two empty circles 
in Fig. \ref{fig1} have been calculated using 
a $N=5000$ droplet (see also Fig. \ref{fig3}). The purpose of these
calculations has been to check explicitly the scaling of 
the results with the number of atoms in the droplet. 
A prolate vortex-free configuration corresponding to $\Lambda =1.6$  is shown in 
Fig. \ref{fig2}.

To allow for a sensible comparison with the experimental\cite{Lan18}
and classical\cite{Bal15} results, we have determined 
the aspect ratio of the superfluid droplets.
For any stationary configuration 
obtained solving Eq. (\ref{eq3}), a sharp density 
surface is defined by calculating
the locus at which the helium density equals $\rho_0/2$; for a
spherical distribution this corresponds to a sphere of radius $R$. 
In the case of deformed droplets, three lengths  
$a, b, c$
are introduced representing the distances from the droplet center 
of mass  to the sharp surface along the 
principal axis of inertia. Following the notation used in the 
experiments and in the classical models, 
we will take $a$ 
and $c$ as the largest and smallest radii, respectively.
For an axisymmetric droplet $a=b \ne c$, whereas $a\ne b \ne c$
in the case of triaxial, prolate shapes.

Figure \ref{fig3}   displays $b^3/V$ {\it vs.} the ratio $a/c$; when  
$a/c=1$, $b^3/V= R^3/V=  3/(4 \pi)$.  
The dashed line and the starred symbols show the classical model 
and the experimental results, respectively, while
the filled dots 
correspond to superfluid-droplet
DFT calculations.
We recall here that all the superfluid droplets corresponding 
to the filled dots 
are prolate.
As for the $\Omega(\Lambda)$ curve in Fig. \ref{fig1}, it may be seen  that the 
rotational  normal fluid and the irrotational superfluid
results are very different for low and intermediate 
values of the angular momentum (corresponding to 
low-to-intermediate values of $a/c$ in Fig. \ref{fig3}), and
they also differ from the experimental results in the same range.
It is worth stressing
here that, at variance with the classical rotating droplet case,
where the rotational axis always coincides with the shortest ($c$)
axis, for superfluid droplets the opposite is true, 
i.e. the rotational axis coincides instead with the intermediate ($b$)
axis.

The discrepancy between our results and the experimental ones
is particularly striking in the region $a/c<1.5$, where 
oblate-like shapes are experimentally found.
We observe at this point that 
oblate-like configurations can only store angular momentum
by nucleating a number of vortices in their interior, 
in the form of a vortex array 
that reduces 
the $D_{\infty h}$ symmetry of the axisymmetric vortex-free droplet 
to, e.g., a $D_{n h}$ symmetry if $n$ vortices are evenly 
distributed in a ring around the droplet center.
Examples of oblate-like configurations hosting a
different number of vortices 
have been presented in Ref. \onlinecite{Anc15}. 

We have thus calculated the equilibrium shapes of rotating $^4$He droplets
hosting vortices
solving also Eq. (\ref{eq3}), this time
using 
the imprinting  procedure\cite{Anc15,Anc17} by which  $n_v$
vortex lines parallel to the $z$ axis
are initially created. This is achieved by starting the  
iterative solution of Eq.(\ref{eq3}) from the effective wave function
\begin{equation}
\Psi_0(\mathbf{r})=\rho_0^{1/2}(\mathbf{r})\, 
\prod _{j=1}^{n_v} \left[ {(x-x_j)+\imath (y-y_j) \over \sqrt{(x-x_j)^2+(y-y_j)^2}}  \right]
\label{eq5}
\end{equation}
where   $(x_j, y_j)$ is the initial position of the $j$-vortex linear  core with
respect to the $z$-axis of the droplet, and $\rho_0(\mathbf{r})$ is the vortex-free 
droplet density profile.
The initial vortex positions are guessed and
during the functional minimization
of the total energy, both the vortex positions and droplet density
are allowed to change
to provide at convergence the lowest energy
configuration for the chosen value  of $\omega$ or $L_z$.

In this case, part of the angular momentum of the droplet is 
stored in vortices 
and part in capillary waves, always present
in prolate configurations. Indeed, for oblate configurations 
one has $L_z \lesssim n_v N \hbar$ with $n_v=1, 2, \cdots$, whereas for
%%%%%%
%  MBG
most
%%%%%%
prolate configurations $L_z > n_v N \hbar$, the extra angular momentum 
being associated to capillary waves.

Figure \ref{fig2} displays one of such prolate configuration hosting 
$n_v=4$ vortices for $\Lambda =1.6$.
This $n_v$ is the largest number of vortices 
the $N=1500$ atoms droplet may accommodate. 
Depending upon the value of $\Lambda $,
the vortex cores can be either arranged in a cross-like 
configuration  (stable at low values of $\Lambda $) or
aligned along the largest droplet axis (stable at high values of $\Lambda $).
Fig. \ref{fig2} also displays 
a prolate droplet hosting four vortices in a cross configuration for $\Lambda =1.05$.\cite{SM}
Notice that the vortex cores meet the droplet surface perpendicularly.

We display in Fig. \ref{fig1} the $\Omega(\Lambda)$ relationship for
the droplets hosting $n_v=3$ and 4 vortices.
The resulting curve is compared 
with the classical model predictions.\cite{Bal15} 
%%%%%%
%FRANC
%%%%%%
Note that, similarly to the classical case, a bifurcation point 
appears, separating the oblate droplets configurations (left branch) 
from the prolate droplets ones (right branch).
The difference between the vortex-hosting droplets case and the 
vortex-free droplets (green dots) shows how different is the way
in which angular momentum is stored in the two cases.

The $b^3/V$ {\it vs.}  $a/c$ curve in the presence of vortices
now displays a rising branch in the left part of the aspect-ratio curve of Fig. \ref{fig3} 
corresponding to  oblate-like configurations,
in agreement with experiments and classical calculations.
Remarkably, the oblate-like part of that curve stops close to the value 
$a/c=1.5$, which is the classical stability limit for axisymmetric configurations.\cite{Cha65} 
Nearly from this $a/c$ value on, the stationary configurations 
have prolate shapes. 
We display in  Fig. \ref{fig3} 
(as well as in Fig. \ref{fig1})
results for both 4-vortex cross and linear configurations 
at $\Lambda$ values where 
they are both stable; according to our calculations, 
cross and linear configurations with the same value of 
$\Lambda$ are nearly degenerate and could both be 
realized in the experiments.
 
\section{Discussion and conclusions}

Figure \ref{fig3} contains the key results of the present study, that we summarize as follows:

$\bullet$
The predicted shapes of vortex-free superfluid droplets 
(green dots in Fig. \ref{fig3}) 
are much different from the 
experimental results\cite{Rup17,Lan18}
in the region of the aspect-ratio chart characterized by low-medium values
of $a/c$ (corresponding to low-medium values of $\Lambda$),
while they agree in the region of higher $a/c$ values (corresponding to 
higher $\Lambda $).

\medskip 
$\bullet$
The presence of vortices drastically modify the sequence of permitted  droplet shapes,
in particular allowing the appearance of stable axisymmetric (oblate-like) shapes,
in agreement with experiments.
Given the way they are produced,
there is no reason why $^4$He drops 
should not host a number of vortices, possibly created  
via the Kibble-Zurek mechanism,\cite{Kib76}
as it happens in liquid $^4$He where 
a fast adiabatic passage through the lambda
transition results in copious vortex production.\cite{Hen93}
%%%%%%%%
%  MBG
%%%%%%%%
%Another possible mechanism for vortex production 
%could be the counterflow heat transfer from the drop center
%to the surface during the cooling process
%down to the droplet final temperature, 
%which could produce turbulence (and thus
%vortices) in the droplet.

\medskip
$\bullet$
Our calculations show that when the number of vortices in the 
droplet is close to its maximum possible (four in the case of the small droplet used in the DFT
calculations), the aspect-ratio $b^3/V$  {\it vs.} $a/c$ curve 
agrees with that obtained using the classical rotating  droplet model.
Some differences are expected to show up  between both approaches, as 
drops in  classical models are considered  incompressible 
and the existence of any surface width is neglected, whereas within DFT
liquid compressibility and surface width are taken into account.
From the overall good agreement between experiments and classical models of 
rotation regarding the droplet morphology,\cite{Rup17,Lan18} one can infer
that  the droplets observed in experiments
host quantized vortices  whose presence could be determined after doping them.\cite{Gom14,Jon16}
Our results also show that vortex arrays are naturally  present  in prolate droplets,
instead of just being  associated with oblate-like  
helium droplets as it is commonly believed. 

\medskip
$\bullet$
Prolate vortex-free  
helium droplets are also stable objects
in which the angular momentum is stored as giant capillary waves  
(like those reported on  charged helium drops levitated with a magnetic 
field\cite{Whi98}). 
Such droplets can achieve much larger  values of $a/c$ than those hosting vortices. In particular,
we have found stable  configurations for $a/c$ as large as 4.5, whereas 
the largest $a/c$ values displayed in Fig. \ref{fig3} for the 
droplets hosting vortices correspond to their actual  stability limit.
According to our results, vortex-free droplets shapes are
indistinguishable from droplets hosting vortices 
unless the rotational axis is identified in the experiments, 
whereas they could be currently detected and distinguished 
from vortex-hosting ones in the range 
$1<a/c\lesssim 1.6$. However, 
they have apparently escaped direct measure so far
(there is only one point with $a/c \sim 1.6$ in the experimental 
aspect ratios that 
could be a potential candidate, see Fig. \ref{fig3}), 
because 
they likely  constitute a fairly 
small fraction of the  --already small-- fraction of  
deformed drops identified in the experiments
(the majority of the imaged $^4$He drops are spherical, 
only  about 1.5 \% of those of Ref. \onlinecite{Lan18}
showed diffraction patterns corresponding to  prolate-like deformed shapes),
making difficult their direct detection.

Finally, we comment briefly on the relative stability of 
configurations having a different number of vortices $n_v$
for a given value of $\Lambda$.
Obviously, only one of the $n_v$ values corresponds to  
the globally stable configuration, the others can only be
metastable. It might be that these configurations are separated by energy barriers whose
existence cannot be determined by the present method, as it only yields the 
lowest energy configuration for a fixed $L_z$ and a chosen $n_v$.
However, according to our calculations, 
these configurations differ in energy by less than 1\%
and it cannot be discarded that they all might show up in the experiments. 
A full analysis of the morphology of the droplets and their 
stability as a function of $n_v$ goes beyond 
the scope of the present work and it is not even necessary for 
%%%%%%%%
%   MBG
%the  discussion of 
discussing the current
%%%%%%%%%%%
experimental results. In fact, 
the calculated structures of droplet hosting vortices
lie along a common aspect ratio curve, as Fig. \ref{fig3} clearly shows,
and even if some of
the calculated points represent metastable states, the main
conclusions of our work will not change.

\bigskip

\begin{acknowledgments}
We thank Jordi Boronat, Andrey Vilesov,  Sam Butler, Thomas M\"oller, and Bruno Langbehn 
for useful discussions and exchanges.  We are most indebted to 
Thomas M\"oller and Bruno Langbehn
for sharing with us their experimental results, and to
Sam Butler for providing us with the
results of  the classical model calculations. This work has been 
performed under Grant No  FIS2017-87801-P from DGI (Spain).
MB thanks the Universit\'e F\'ed\'erale Toulouse 
Midi-Pyr\'en\'ees for financial support  throughout the ``Chaires d'Attractivit\'e 2014''  Programme IMDYNHE. 
\end{acknowledgments}


\begin{thebibliography}{99}
\bibitem{note}
We use the  $\Lambda$ definition of Refs. \cite{Hei06,But11,Ber17} that is four  times larger than that of 
Refs. \cite{Bro80,Bro80b}.
\end{thebibliography}
\end{document}